\documentclass[12pt,a4paper]{article}
\usepackage{graphicx}

\author{O.I. Berngardt}

\begin{document}

%
%

\title{Seismo-ionospheric effects associated with 'Chelyabinsk' meteorite during the first 25 minutes after its fall}

%
%

%
%

\author{O.I. Berngardt}









%
%


\maketitle

\begin{abstract}
This paper presents the properties of ionospheric irregularities elongated
with Earth magnetic field during the first 25 minutes after the fall
of the meteorite 'Chelyabinsk' experimentally observed with EKB radar
of Russian segment of the SuperDARN. 

It is shown that 40 minutes before meteor fall the EKB radar started
to observe powerful scattering from irregularities elongated with
the Earth magnetic field in the F-layer. Scattering was observed for
80 minutes and stopped 40 minutes after the meteorite fall. 

During 9-15 minutes after the meteorite fall at ranges 400-1200 km
from the explosion site a changes were observed in the spectral and
amplitude characteristics of the scattered signal. This features were
the sharp increase in the Doppler frequency shift of the scattered
signal corresponding to the Doppler velocities about 600 m/s and the
sharp increase of the scattered signal amplitude. This allows us to
conclude that we detected the growth of small-scale ionospheric irregularities
elongated with the Earth magnetic field at E-layer heights. Joint
analysis with the seismic data and numerical modeling shows that the
observed effect is connected with the passage of secondary acoustic
front formed by supersonic seismic ground wave from the 'Chelyabinsk'
meteorite. 

As a possible explanation the growth of elongated ionospheric irregularities
may be caused by the passage of the high-speed acoustic wave in the
ionosphere in the presence of high enough background electric field.

\end{abstract}

%
%

%


%
%

\section{Introduction}

The fall of 'Chelyabinsk' meteorite 03:20UT 15/02/2013 was accompanied
by a large number of dynamic ionospheric \cite{Tertyshnikov_et_al_2013, Givishvili_et_al_2013, Gohberg_et_al_2013, Berngardt_et_al_2013, Yang_et_al_2014},
atmospheric \cite{LePichon_et_al_2013, Gorkavyi_et_al_2013} and seismic
\cite{Tauzin_et_al_2013} phenomena. The well known dynamic ionospheric
effects include formation of midscale radially traveling ionospheric
disturbances (MSTIDs) at F-layer heights, detected directly by EKB
coherent radar and GPS network \cite{Berngardt_et_al_2013b, Gohberg_et_al_2013, Yang_et_al_2014}
(at distances less than 1000 km). Also an indirect evidence of the
same radial waves at E-layer heights \cite{Kutelev_and_Berngardt_2013}
(at distances less than 700 km) were found. The analysis shows that
the average speed of this MSTIDs, observed at 500-1000 km distances
from the epicenter, is 400m/s and 200 m/s\cite{Kutelev_and_Berngardt_2013}.
Analysis of GPS data provides estimates of the average velocity of
140 m/s\cite{Berngardt_et_al_2013}, 400 m/s \cite{Berngardt_et_al_2013b,Yang_et_al_2014}
and 800 m/s \cite{Yang_et_al_2014}. Basic wave effects thus appear
between 20-th and 80-th minutes after the explosion. 

Significant ionospheric effects during the first 20-25 minutes after
the fall at close distances from the explosion site (about 500-700km)
were not investigated yet.

In this paper we studied the ionospheric effects observed in the first
25 minutes after the meteorite fall in E- and F-layers of the ionosphere.
The nearest zone to the explosion site (<500-700km) is potentially
the most disturbed. Therefore the study of ionospheric effects in
this region requires high spatial and temporal resolutions simultaneously,
which is currently provided only by EKB radar.

The EKB radar is the first coherent decameter radar of Russian segment
of SuperDARN \cite{Chisham_et_al_2007}. It was bought, assembled
and started by ISTP SB RAS for monitoring in the middle of December
2012. This allowed us to get a big amount of the ionospheric data during
the meteorite fall, as well as before and after the fall with high
spatially-temporary resolution. EKB radar is located around 200 km
to north-west from the meteorite explosion site. The radar is the
analog of CUTLASS radar \cite{Lester_et_al_2004} positioned by ISTP
SB RAS at  'Arti'  observatory of Institute of Geophysics, Ural Branch
of Russian Academy of Sciences (56o26 N, 58o34 E).

The basis of the radar functionality is monitoring of back-scattered
signal characteristics simultaneously in back-oblique sounding mode
(BOS, ground backscatter) and the backscattering from small-scale
irregularities mode (BS). This allows to make simultaneous estimation
of both the characteristics of the background ionosphere in BOS mode,
and the characteristics of small-scale irregularities in BS mode.
The antenna system of the radar is a phased array with azimuthal scan
sector of about 50 degrees and with a mainlobe azimuthal width of
about 3 degrees. Scanning the entire sector is made by iterating over
16 directions counterclockwise. The scanning takes for about 60 seconds,
with probing in each of the 16 fixed beam directions for about 4 seconds. 

\begin{figure}
\includegraphics[scale=0.7]{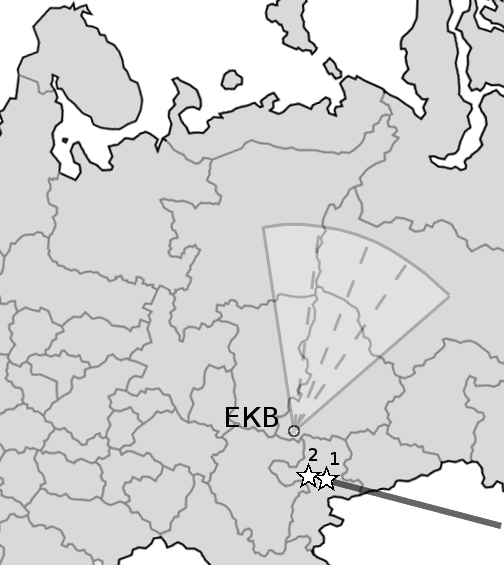}
\caption{EKB radar field-of-view and trajectory of meteorite 'Chelyabinsk',
its explosion (1) and fall(2).
}
\label{fig:1} 
\end{figure}

During the observations the radar worked with 60km range resolution
in the range of distances 400-3500km. The area of a meteorite explosion
and fall was located 200km to the south of the radar in the area of
the antenna pattern backlobe. The geometry of the experiment is shown
at Fig.\ref{fig:1}. Further in the paper we will refer the time and place of
maximum luminosity 03:20:33 UT as the moment and site of the explosion,
and refer the place of finding the main meteorite fragment (Chebyrkul
Lake) as the place of the fall.

Radar EKB has a unique spatial and temporal resolution, allowing detailed
study of many ionospheric effects during the meteorite 'Chelyabinsk'
fall. Some of the ionospheric effects based on the radar data are
described in \cite{Berngardt_et_al_2013,Berngardt_et_al_2013b,Kutelev_and_Berngardt_2013}.
All these papers are focused to an analysis of the effects in the
electron density profile during the first 2 hours after the fall.
These effects mainly related with the changes in the background electron
density with the spatial scales exceeding the size of the Fresnel
zone.

In this paper we will study the effects of amplifying short-lived
small-scale irregularities with hundred-meter spatial scales, smaller
than the Fresnel radius, that hardly be detected by other ionospheric
instruments.

\section{Ionospheric observations}

According to the literature \cite{Berngardt_et_al_2013,Berngardt_et_al_2013b,Alpatov_et_al_2013,Givishvili_et_al_2013,Gohberg_et_al_2013,Tertyshnikov_et_al_2013}
the meteorite fall was characterized by a relatively quiet geomagnetic
and seismic conditions, as well as the absence of significant solar
flares. Except for regular disturbances associated with a passage
of the solar terminator, and with the impact of the previous day 14/02/2013,
characterized by weak geomagnetic disturbance the average ionospheric
situation was quiet. Therefore, ionospheric dynamics should be similar
to the dynamics of the nearest quiet days. This allow us to differ
the effects associated with the explosion and the fall of the meteorite
from a regular daily dynamics with a high degree of confidence. To
highlight effects of irregular dynamics in the power of the scattered
signal, associated with regular processes in the ionosphere, we conducted
an analysis of 15/02/2013 data relative to the same quiet (the reference)
days 9-12,18 / 02/2013, following to \cite{Berngardt_et_al_2013b}.

\begin{figure}
\includegraphics[scale=0.72]{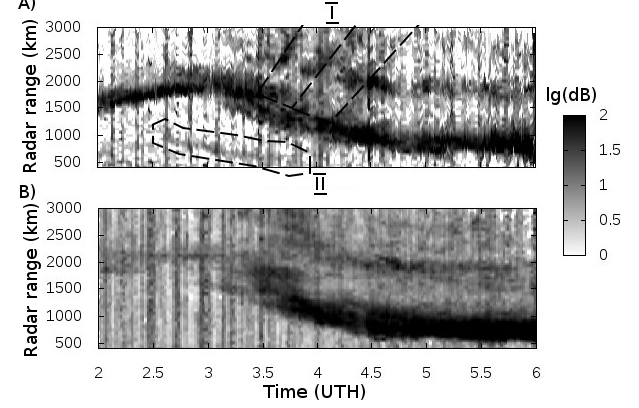}

\caption{Average (over the field-of-view azimuths) received signal power
from EKB radar, as a function of radar range and time. A) - during
the day of the meteorite fall 15/02/2013; B) - averaged over the reference
days 9-12,18/02/2013.
}
\label{fig:2} 
\end{figure}

The main technique for detection of irregular effects is to analyze
power of the received signal 15/02/2013, and to compare it with the
power averaged over the the reference days. For qualitative analysis
the scattered signal power were averaged over the entire field-of-view
azimuths as a function of time and distance. To simplify the qualitative
analysis we consider only the cases of high level of scattered signal
power exceeding the noise level. Fig.\ref{fig:2}A shows the overall picture
of the average power of the scattered signal, averaged over the field-of-view
azimuths during 15/02/2013, and Fig.\ref{fig:2}B - the same but averaged over
the reference days. Region I studied in detail in \cite{Kutelev_and_Berngardt_2013}.
In this paper we will study in detail the region II, associated with
the effect of scattering from small-scale F-layer irregularities during
the period 02:45-04:00UT.

According to EKB radar data, 40 minutes before the meteorite fall
a signal was observed coming from the ranges 500-1000km (Fig.\ref{fig:2}A, region
II). Earlier, in the work \cite{Berngardt_et_al_2013b} it was suggested
that a possible reason for the formation of such a signal was increase
of the electron density in the E-layer of the ionosphere. Detailed
analysis, however, showed that the spectral width of the signal often
exceeds 100-200 m/s, and the signal itself is observed at a distance
equal to half the distance the ground backscatter signal. This suggests
that the cause of the signal between 02:45-04:00UT is the scattering
by F-layer irregularities elongated with the Earth magnetic field.

The main mechanism of formation of these irregularities are usually
considered plasma instabilities, particle precipitation and turbulence
in the ionospheric plasma \cite{Fejer_and_Kelley_1980}. The growth
of such irregularities, as a rule, requires a sufficiently high electric
field at ionospheric heights, that usually associated with increased
activity in the near-Earth space. 

It should be noted that when the background electric field exceed
25mV/m, there are not only elongated irregularities in the F-layer
grown, but also irregularities in the E-layer are grown, which are
associated with Farley-Buneman instabilities \cite{Farley1963,Buneman1963}.
Such irregularities in E-layer in a simple manner can be separated
from the F-layer irregularities because they often observed at a fixed
radar range, not dependent on the refraction effects, that are typical
for radio signals propagation in the F-layer. Since position of the
region II depend on time, as it predicted by refraction effect, the
region II corresponds to F-layer scattering. So E-layer irregularities
has not been observed before meteorite fall (02:40-03:20UT) and it
can be assumed that the background electric field during 02:40-03:20UT
was strong enough to generate irregularities in the F-layer, but did
not exceed 25mV/m necessary for the generation of E-layer irregularities.

\begin{figure}
\includegraphics[scale=0.6]{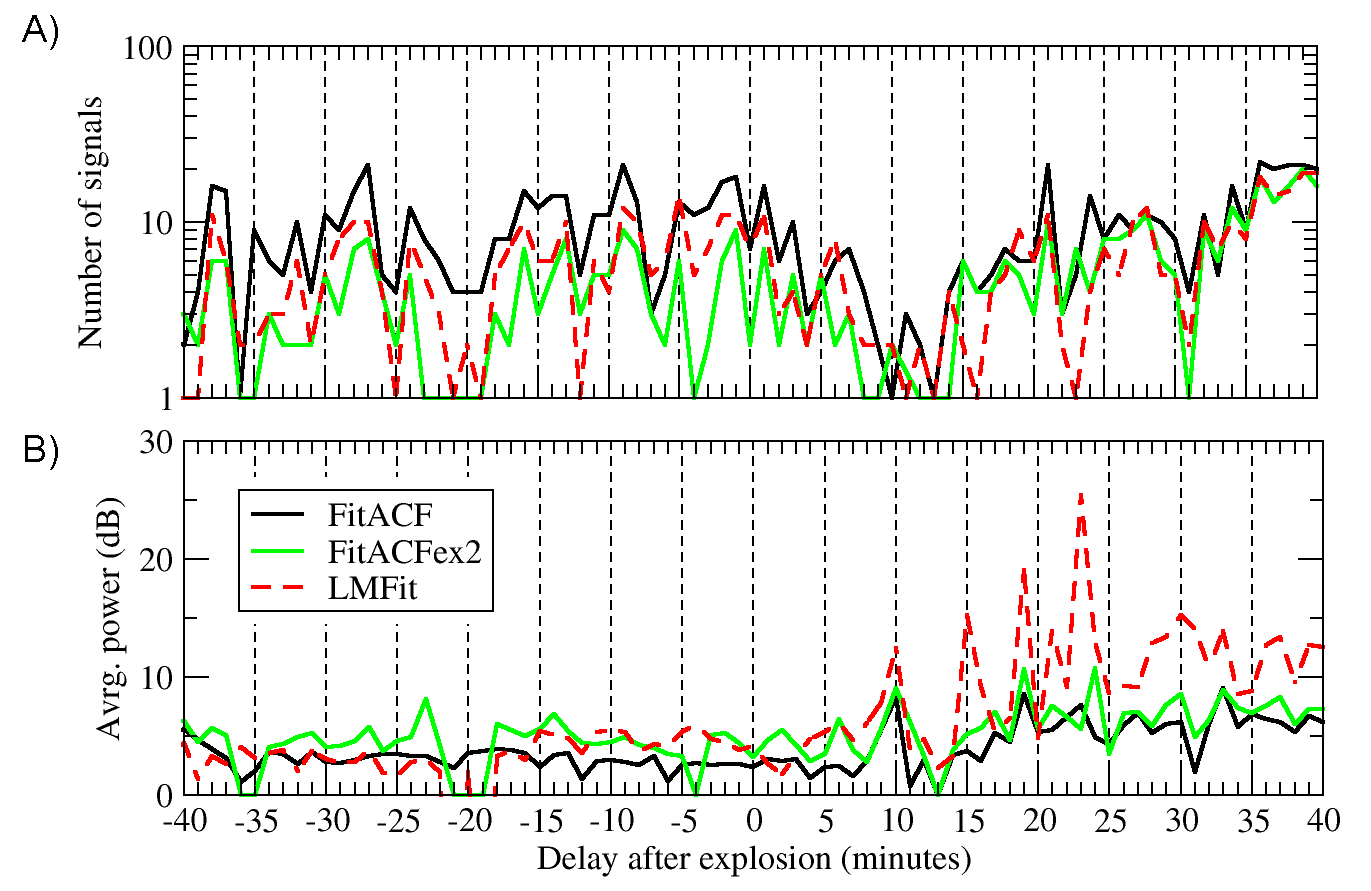}
\caption{The result of statistical analysis of the EKB radar data in
the range of distances 500-1200km in times close to the time of the
explosion. Calculations were carried out by three algorithms - FitACF,
FitEx2 and LMFit. A) - the amount of 'good' data per 1 minute that
can be processed B) - average power of the signal calculated from
the 'good' data.
}
\label{fig:5}
\end{figure}

Let us analyze in detail the behavior of the scattered signal at EKB
radar during +/- 40 minutes of the explosion. Currently, for the estimating
of the parameters of ionospheric irregularities from the SuperDARN
scattered signal the three basic algorithms are used - FitACF, FitACFex2
and LMFit \cite{Ribeiro_et_al_2013}. Fig.\ref{fig:5} shows the result of analysis
by all three programs during 15/02/2013. Fig.\ref{fig:5}A shows the number of
'good' data per 1 minute at all lags corresponding to the distances
500-1200km from the radar, at which the scattering from elongated
E-layer irregularities can be observed. The 'good' data means that
result of calculations gives physically meaningful parameters - spectral
width greater than 0, and signal-to-noise ratio greater than 0dB.
The average power of the scattered signal calculated from the good
data, is shown at Fig.\ref{fig:5}B.

\begin{figure}
\includegraphics[scale=0.8]{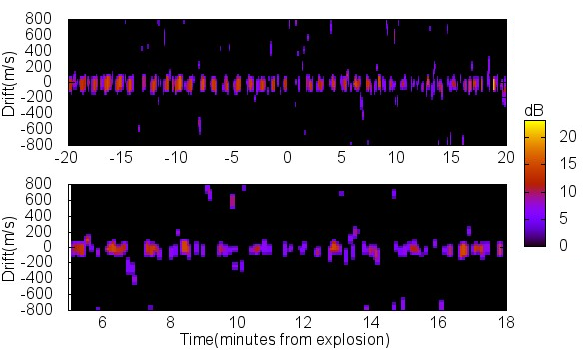}
\caption{Mean spectral shape of scattered signal in the range 800-1200km from the radar, 
as a function of delay from explosion moment. Frequency is given in Doppler drift (m/s) units.
}
\label{fig:5bis}
\end{figure}

Complex structure of the spectra is shown at Fig.\ref{fig:5bis}. 
The spectra obtained by integration over the radar ranges 800-1200km and 
subtracting mean noise. 
One can see that during 9-15 min. after the meteorite explosion spectra 
zero-drift becomes narrower, and spectra itself becomes more complex, 
having up to 3-peaks, that explains difficulty in theirs interpretation 
by standard programs. One can see, that most powerful and high-speed response
is observed at 9th-10th minutes after the meteorite fall. Less powerful
effects are also observed at 7-th minute.

It can be seen from the Fig.\ref{fig:5} that during the period from 7 to 17
minutes after the meteorite explosion all the algorithms show a sharp
decrease in the number of 'good' data. The results of processing by
the most recent and accurate FitACFex2 and LMFit programs \cite{Ribeiro_et_al_2013}
show a dramatic increase in the power of the scattered signal at 10,
15, 19 and 23-24 minutes after explosion. A sharp increase in the
signal power at 10 and 19 minutes can also be seen from FitACF program
results. Monotonic increase in signal power after 15 minute is caused
by the regular ionospheric processes in the dawn time.

This suggests that in the period from 10 to 17 minute the scattered
signal becomes quite complicated and can not be accurately processed
by standard techniques based on the assumptions of FitACF,FitACFex2
and LMFit models, and can be analyzed only qualitatively. At the same
time, it is obvious that at 10, 15, 19 and 23 minutes the amplitude
of the scattered signal sharply increased, which indicates the possible
growth of small-scale scattering irregularities during these periods.

The main mechanism for the formation of the E-layer irregularities
at mid-latitudes is usually the growth of two-stream instabilities
\cite{Farley1963,Buneman1963}, usually classified as Type I scattering,
characterized by high drift velocities equal to or greater than the
ion-acoustic speed in the ionosphere.

\begin{figure}
\includegraphics[scale=0.5]{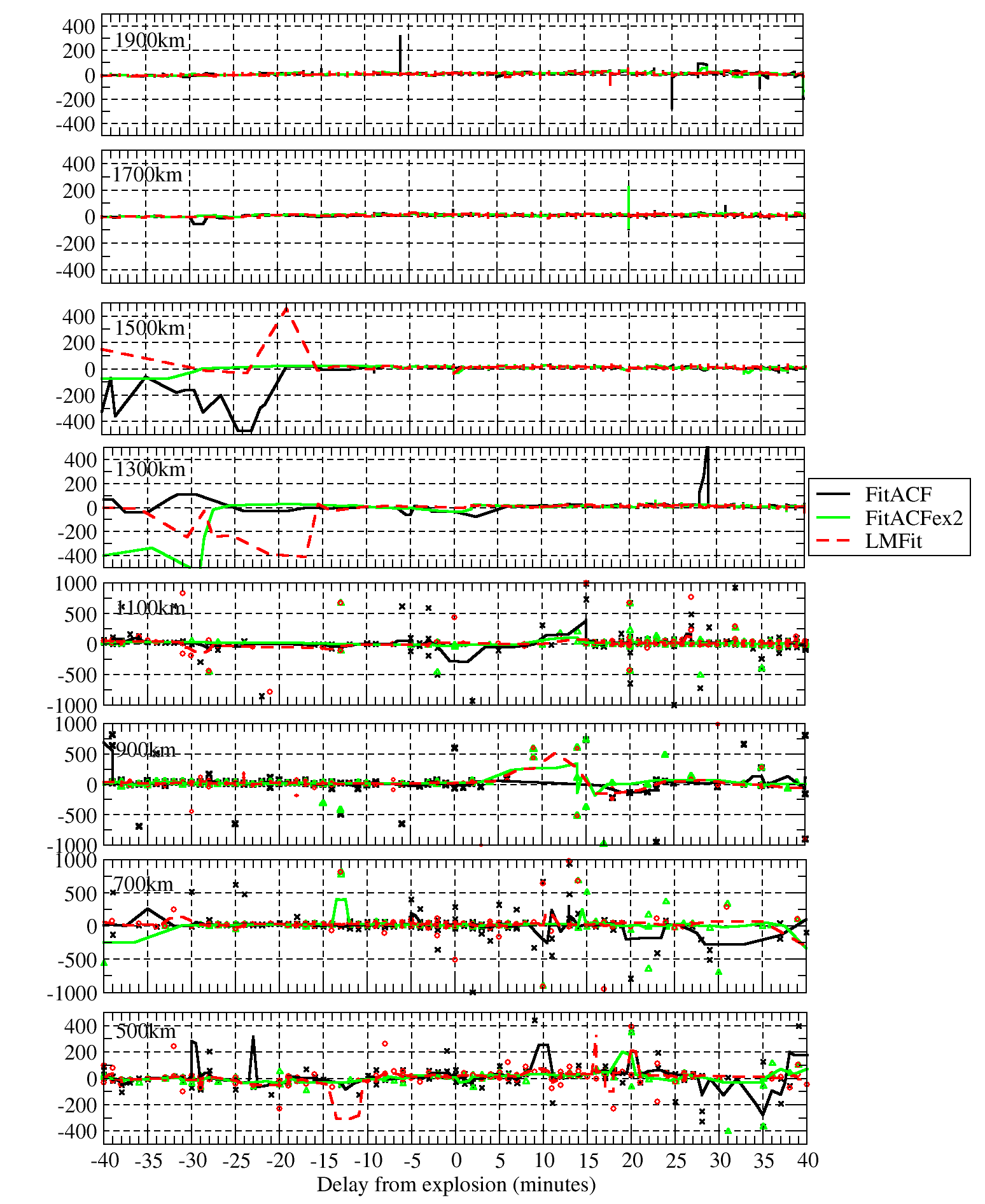}
\caption{Doppler velocity, calculated from the scattered signal, as
a function of delay after the explosion and distance from the EKB
radar, calculated by FitACF, FitACFex2 and LMFit programs. Circles,
triangles and crosses represent the calculated values , lines - their
medians over the 4 adjacent values
}
\label{fig:6}
\end{figure}

Fig.\ref{fig:6} shows the values of the Doppler velocity computed by the scattered
signal as a function of delay after the explosion and distance from
the radar. Calculations were carried out by three different programs:
FitACF, FitACFex2 and LMFit. Circles, triangles and crosses represent
the values computed by algorithms, lines - the medians over the 4
adjacent values, allowing to get rid of the occasional mistakes and
get smoothed values. From Fig.\ref{fig:6} it is seen that 
all three algorithms show an increase in Doppler velocity at ranges 
500-900km during 7th-14th minutes after the explosion, with an absolute 
maximum at 9th-10th minute.

As it has been shown (Fig.\ref{fig:5bis}) the scattered signal at this time is
quite complicated, and the parameters of the scattered signal are
harder to determine than usually. Nevertheless, the fact that different
methods shows similar high mean values of Doppler shift, allows to
suggest that at 10-th minute after the explosion at ranges 500-1200km
from the epicenter at EKB radar we actually observed scattered signal
with high Doppler shifts, corresponding to the scattering from elongated
E-layer irregularities of Type I (Farley-Buneman).

\section{Seismo-ionospheric processes}

\begin{figure}
\includegraphics[scale=0.6]{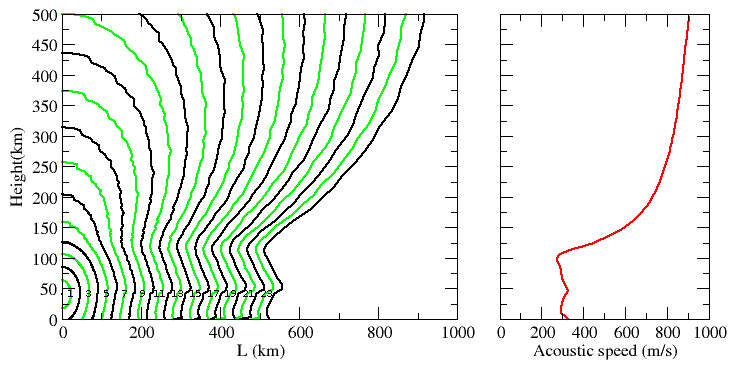}

\caption{ A) - The model of propagational delay of the acoustic signal
at different heights and distances from the source, measured across
the meteorite trajectory (source height 41km), contour lines every
1 minute; B) - the profile of sound speed in the ionosphere as a function
of height, used for modeling.
}
\label{fig:7}
\end{figure}

Numerical simulation of the acoustic signal in the ionosphere, from
the source at 41km altitude, corresponding to the first intense flare
of the bolide \cite{Borovicka_et_al_2013} is shown at Fig.\ref{fig:7}. The
simulation showed that the delay of the acoustic signal propagating
directly over the path meteor-observation point at 250 km height (for
the minimum observed radar range - 500 km) is 17-18 minutes, and at
the height 90-120km for the same distance - at least 23 minutes. So
the propagation of this 'primary' acoustic wave from the meteorite
trajectory can not explain the effects at ranges 500-1200km, observed
with a delay of less than 17 minutes from explosion.

Known effect of the flyover and the explosion of the meteorite is
the formation of acoustic \cite{Popova_et_al_2013,LePichon_et_al_2013}
and seismic \cite{Tauzin_et_al_2013} waves. Seismic waves observed
during meteorite falls, are usually caused by three reasons - by striking
massive body to the earth's surface, by meteorite explosion and by
coming the acoustic wave produced by the passage of a supersonic body
in the upper atmosphere. Review of techniques for the recovery of
the trajectory of a meteorite based on these effects is shown, for
example in \cite{Edwards_et_al_2008}.

\begin{figure}
\includegraphics[scale=0.5]{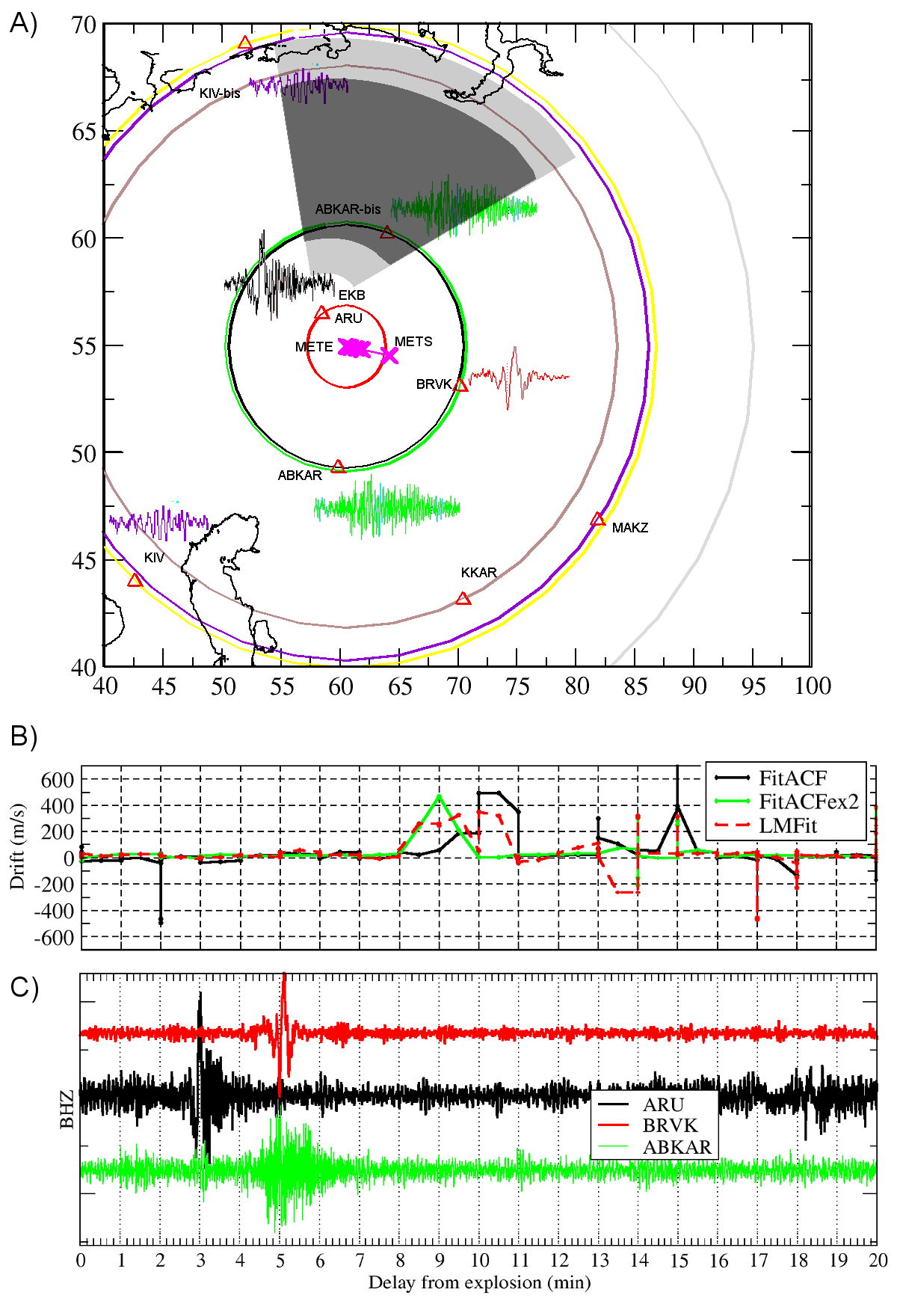}
\caption{ A) - The geometry of seismic observations, B) - Doppler velocities
of ionospheric irregularities according to radar EKB, computed by
three algorithms - FitACF, FitACFex2 and LMFit, as a function of delay
after the meteorite explosion, and C) - vertical seismic oscillations
as a function of delay after the meteorite explosion. ABKAR-bis and
KIV-bis - points that are symmetrical to observatories ABKAR and KIV
relative to the meteorite trajectory. Light gray sector shows field-of-view
of the EKB radar, the dark gray sector - the range of distances that
meets conditions for observation of scattering from the E-layer irregularities.
}
\label{fig:8}
\end{figure}

The mass of the 'Chelyabinsk' meteorite fragments was relatively small,
and the fall of the main fragment happened into the water, so the
impact mechanism of seismic waves, apparently, less powerful than
the acoustic from explosion and flyover. This can be approved by analysis
of seismic data from seismic stations close to the crash site - ARU,
ABKAR, BRVK (Fig.\ref{fig:8}A,C).

The main seismic signal came to the ARU station 3 minutes after the
explosion (200 km from the epicenter, the equivalent ground speed
1.1km/s), and to the stations ABKAR and BRVK - 5 minutes after the
explosion (600 km from the epicenter, the equivalent ground speed
2km/s). A similar effect of changing the apparent velocity as a function
of distance from the crash site was observed in the paper \cite{Tauzin_et_al_2013}.
Therefore, if considering the mechanism of formation of the seismic
signal due to the passage of a supersonic body, for the calculation
of the horizontal velocity it must be taken into account the delay
due to sound propagation from height of flight (30-60km) to the ground
(\textasciitilde{} 2 min, Fig.\ref{fig:8}). Taking into account the delay for
the calculation of the seismic disturbance ground speed results to
the estimates of the seismic waves speed 3.3km/s, known for large
distances \cite{Tauzin_et_al_2013}, the same for all three seismic
stations (ARU, ABKAR, BRVK). This indicates good validity of acoustic
model for the formation of the main source of the seismic signal in
this case.

In this model, the source of the acoustic signal is not the point
of the fall, but the entire flight path of the meteorite \cite{Edwards_et_al_2008},
and the wave itself can be considered axisymmetric, with the front
close to the cylinder (at small distances from the epicenter). In
this case, we can assume that the seismic disturbance signals have
axial symmetry, and the signal at the point ABKAR-bis (its place corresponds
to symmetric reflection of position of ABKAR seismic observatory relative
to the path of the meteorite) (Fig.\ref{fig:8}A), is expected to be similar
to the signal at the point ABKAR.

Usually it is assumed that the fine structure of the signal (its main
spectral components) is significantly affected by Earth topography
and characteristics at the measurement point \cite{Edwards_et_al_2008},
but the envelope of the seismic signal and the group delay should
correlate reasonably well at the same distances and axial angles to
the meteorite trajectory, which allows to estimate with good accuracy
the delay of the seismic signal at ABKAR-bis point, located almost
under the EKB-radar observation area and consider it repeating ABKAR
seismic signal.

By estimating acoustic velocity from the MSIS model temperature profile
(Fig.\ref{fig:7}.B), we can estimate the propagation time of acoustic disturbance
caused by the propagation of the supersonic seismic wave from the
ground to the E-layer heights (100 km, 5 minutes). Due to the high
speed of seismic wave the propagation direction of the acoustic wave
(determined by the relation between sound speed and the speed of the
source) is almost vertical. Comparison of calculated delays with experimental
observations (Fig.\ref{fig:8}B, C) shows good consistency. Limited range of the
area where observation of scattering is possible in E-layer (Figure
6) is also consistent with the model.

The effect of propagating seismic wave causes the formation of inhomogeneities
in the ionospheric plasma, that are experimentally observed after
the earthquakes \cite{Maruyama_et_al_2012} and has a theoretical
explanation \cite{Maruyama_and_Shinagawa_2014}. It should be noted
that the vertical amplitude of the seismic waves observed after the
earthquake Tohoku about 300 times higher than from the meteorite 'Chelyabinsk'.
For the linear mechanism of formation of acoustic and ionospheric
irregularities it allows us to expect the amplitude of the perturbations
of the electron density 300 times smaller than for perturbations accompanying
Tohoku earthquake. Characteristic vertical scale inhomogeneities,
calculated from the period of the seismic variations at the station
ABKAR and sound speed in the ionosphere, is about 600-1000 meters.

\begin{figure}
\includegraphics[scale=1.0]{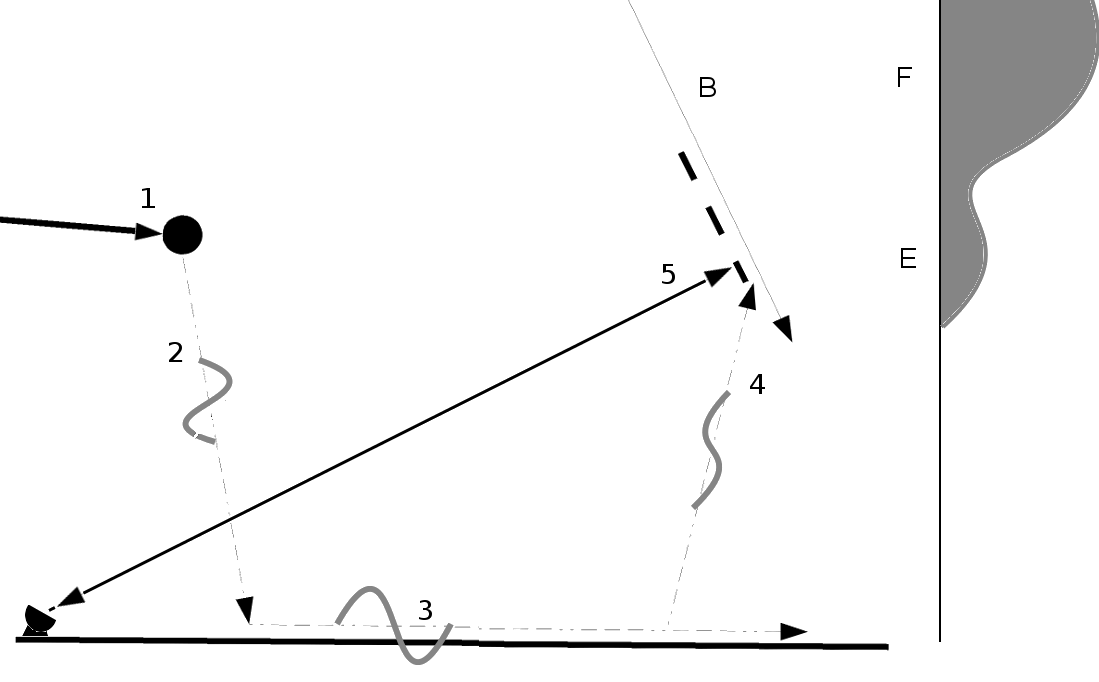}
\caption{ The proposed scheme to explain the observations with the EKB
radar. 1-trajectory of a meteorite; 2-primary acoustic wave; 3-seismic
wave; 4 - secondary acoustic wave; 5- field-aligned scattering of
the EKB radar signal at E-layer heights.
}
\label{fig:9}
\end{figure}

At Fig.\ref{fig:9} it is shown suggested mechanism of the observed effect that
explains delays between different observations.

According to our proposed interpretation of the observed effects in
the E-layer of the ionosphere in the first 25 minutes after the meteorite
explosion, the effect is a superposition of known or previously observed
effects (Fig.\ref{fig:9}).

Acoustic signal caused by the flyover of the meteorite in the upper
atmosphere, after 2 minutes of flight and explosion formed a seismic
source on the ground, which elongated with the projection of the meteorite's
trajectory, by analogy with \cite{Edwards_et_al_2008}. Seismic waves
propagate from the source with speed of about 3.3 km/s \cite{Tauzin_et_al_2013},
forming in the process of their propagation almost horizontal front
of acoustic wave that propagates almost vertically, by analogy with
\cite{Maruyama_et_al_2012}. As a result of interaction of acoustic
wave with the ionized component a quasi-periodic inhomogeneity of
the electron density is formed with spatial scales defined by period
of seismic waves and vertical velocity of the acoustic wave in the
ionosphere, by analogy with \cite{Maruyama_and_Shinagawa_2014}. When
the seismic signal reaches the range 400-600km (4-5 min after the
explosion of a meteorite) it is observed by seismic observatories
ABKAR and BRVK. According to the acoustic wave propagation model,
the delay between observations of vertical ionospheric irregularities
due to the passage of an acoustic signal at E-layer heights generated
by seismic source and the seismic signal under this ionospheric point
is 4-5 minutes.

According to our interpretation the effect of sharp increase in Doppler
velocity of irregularities elongated with the magnetic field, detected
by EKB radar at the 9-11 minute after the meteorite explosion may
be associated with additional growth of the irregularities elongated
with the Earth magnetic field due to the passage of the secondary acoustic
wave through the ionospheric plasma (or corresponding quasi-vertical
traveling of ionospheric plasma inhomogeneties) in the presence of
strong enough background ionospheric electric field. The distance
to the bursts in Doppler velocity is determined by the implementation
of the necessary conditions for scattering from field-aligned irregularities
in E-layer. Time of occurrence of the bursts (delay from the meteorite
explosion time) is associated with the passage of an acoustic wave
through the region of the effective aspect scattering.

The mechanism of such amplification of field-aligned E-layer irregularities
in the presence of an acoustic wave at the moment is not clear. Nevertheless,
as possible growth mechanisms for these elongated irregularities may
be used the following models or theirs combination: the electric field
amplification in the presence of heavy ions in sporadic-E and acoustic
wave \cite{Liperovsky_et_al_1997}; the growth of gradient-drift instability
\cite{Rogister_and_DAngelo_1970} in the presence of vertical gradients
in the electron density caused by the propagation of acoustic waves;
the growth of Kelvin-Helmholtz instability and large electric fields
in E-layer \cite{Bernhardt_2002} in the presence of fast propagating
acoustic wave front.

\section{Conclusion}

This paper presents the properties of ionospheric irregularities elongated
with Earth magnetic field during the first 25 minutes after the fall
of the meteorite 'Chelyabinsk' experimentally observed with EKB radar
of Russian segment of the SuperDARN. 

It is shown that 40 minutes before meteor fall the EKB radar started
to observe powerful scattering from irregularities elongated with
the Earth magnetic field in the F-layer. Scattering was observed for
80 minutes and stopped 40 minutes after the meteorite fall (Fig.\ref{fig:1}).
Most probably this effect is caused by presence of auxiliary electric field.

During 9-15 minutes after the meteorite fall at ranges 400-1200 km
from the explosion site a changes were observed in the spectral and
amplitude characteristics of the scattered signal. This features were
the sharp increase in the Doppler frequency shift of the scattered
signal corresponding to the Doppler velocities about 600 m/s (Fig.\ref{fig:7})
and the sharp increase of the scattered signal amplitude(Fig.\ref{fig:6}). This
feature was detected in the data processed by the three currently
used methods of SuperDARN signals - FitACF, FitACFex2 and LMFit. This
allows us to conclude that we detected the growth of smallscale ionospheric
irregularities elongated with the Earth magnetic field at E-layer
heights.

Joint analysis with the seismic data and numerical modeling shows
that the observed effect is connected with the passage of secondary
acoustic front formed by supersonic seismic ground wave from the 'Chelyabinsk'
meteorite in the presence of a sufficiently high background electric
field. The secondary acoustic wave is formed a supersonic seismic
ground wave caused by the explosion of a meteorite and its passage.
Under this explanation, the distance to bursts of Doppler velocity
is determined by the aspect sensitivity of scattering process in
the E-layer and the geometry of the Earth magnetic field. Moments
of the bursts in Doppler shift (delay them from the meteorite explosion
moment) are associated with the passage of an acoustic wave through
the region of the effective aspect scattering.

As a possible explanation the growth of elongated ionospheric irregularities
may be caused by the passage of the high-speed acoustic wave in the
ionosphere in the presence of high enough background electric field.
The mechanism of such gain of field-aligned E-layer ionospheric irregularities
at the moment is not clear and requires detailed investigation.


%
%
%
%
%
%
%

\section*{Acknowledgments}

This work was supported by the RFBR grant No 14-05-00514a. 
We are grateful to IRIS network for providing data of seismic 
observatories ARU, ABKAR and BRVK. The authors are grateful 
A.Lyahov and T.Loseva (IDG RAS) for useful discussions.

\newpage

\begin{table}

\begin{tabular}{|c|c|c|c|c|}
\hline 
Point name & N & E & Distance & Type \\
\hline 
\hline 
ABKAR & 49.256 & 59.943 & 630km & Seismic station \\
\hline 
BRVK & 53.058 & 70.283 & 670km & Seismic station \\
\hline 
ARU & 56.430 & 58.562 & 220km & Seismic station \\
\hline 
EKB & 56.430 & 58.562 & 220km & Coherent decameter radar \\
\hline 
METS & 54.508 & 64.266  & 240km & Trajectory start, h=91km \\
\hline 
METE & 54.922 & 60.606  & 0km & Trajectory end, h=15km \\
\hline 

\label{tab:1} 

\end{tabular}

\caption{Coordinates of main instruments and points used for data analysis}

\end{table}

\newpage

\newpage

\newpage

\newpage

\newpage

\newpage

\newpage

\end{document}